\documentclass[12pt]{article}
\usepackage[utf8]{inputenc}
\usepackage[pdftex]{graphicx}
\usepackage{hyperref}
\usepackage{amsfonts,amssymb,amsmath}

\newtheorem{definition}{Definition}
\newtheorem{example}{Example}
\newtheorem{lemma}{Lemma}
\newtheorem{theorem}{Theorem}
\newtheorem{remark}{Remark}

\title{Modified L\'evy Laplacian on manifold and Yang--Mills instantons}
\author{Boris O. Volkov$^{1}$}

\begin{document}

\maketitle

$^1$ Steklov Mathematical Institute,  ul. Gubkina, 8, Moscow, 119991, Russia
\\
E-mail: borisvolkov1986@gmail.com

\bigskip

Dedicated to Prof. I.~V.~Volovich  on his 75th anniversary.

\begin{abstract}
An infinite dimensional Laplacian defined as the Ces\'aro mean of the second order directional derivatives on manifold is considered.
This Laplacian is  parameterized  by the choice of a curve in the group of orthogonal rotations. It is shown that, under certain conditions on the curve, this operator is related to instantons on a 4-dimensional manifold.
\end{abstract}

keywords: L\'{e}vy Laplacian; Yang--Mills equations; instantons.

\section*{Introduction}

The L\'evy Laplacian is an infinite dimensional  Laplacian defined as the Cesaro mean of the second order directional derivatives (see~\cite{L1951}).
Let $E\subset H\subset E^\ast$ be a Gelfand triplet. Here $H$ is a real separable Hilbert space and $E$ is a locally convex space continuously embedded into $H$. Let $\{e_n\}$ be an orthonormal basis in $H$ consisting of elements from
$E$. Let $f$ be two times Frechet differentiable function on $E$. Then the value of the L\'evy Laplacian $\Delta_L^{\{e_n\}}$
on the function $f$ is
\begin{equation}
\label{Levy2}
\Delta_L^{\{e_n\}}f(x)=\lim_{n\to\infty}\frac 1n \sum_{k=1}^n\langle f''(x) e_k,e_k\rangle.
\end{equation}
Thus, the definition of the L\'evy Laplacian depends on  the choice of a Hilbert space $H$
and an orthonormal basis $\{e_n\}$.

It turns out that this operator is related to the Yang--Mills gauge fields. Accardi, Gibilisco and  Volovich proved the following assertion (see~\cite{AGV1994}).  A connection  $A$ in a vector bundle over $\mathbb{R}^d$ satisfies the Yang--Mills equations 
$$
D_A^\ast F=0
$$
if and only if
a parallel transport $U^A$ associated with the connection $A$ satisfies the Laplace equation for the L\'evy Laplacian $\Delta_L$:
$$
\Delta_LU^A=0.
$$
Here $F=dA+A\wedge A$ is the curvature  and $D_A^\ast$ is adjoint operator to the covariant exterior derivative.
Note the L\'evy Laplacian in~\cite{AGV1994} was defined in a 
 different way from~(\ref{Levy2}), namely as a special integral functional defined by a special form of the second derivative.
But it was proved in~\cite{Volkov2012,Volkov2020}  that this L\'evy Laplacian can be defined by~(\ref{Levy2}) if we choose the Sobolev space $H^1([0,1],\mathbb{R}^d)$ as $E$, the Hilbert space $L_2([0,1],\mathbb{R}^d)$ as $H$ and some natural basis $\{e_n\}$ in $L_2([0,1],\mathbb{R}^d)$ (see also~\cite{Volkov2021}). The theorem on equivalence of the Laplace equation for the  L\'evy Laplacian and the Yang--Mills equations  was generalized for manifolds (see~\cite{LV2001,Volkov2021}).

It is a natural question to ask if the L\'evy Laplacian related to instantons and antiinstantons on 4-manifolds (see~~\cite{Accardi}), i.e. solutions of the self-duality Yang--Mills equations
$$
F=-\ast F
$$
or the anti self-duality Yang--Mills equations
$$
F=\ast F.
$$ 
It turns out that the L\'evy Laplacian introduced in~\cite{AGV1994} is not invariant under the action of infinite dimensional rotations $W\in C^1([0,1],\mathrm{SO}(4))$.
So any rotation $W\in C^1([0,1],\mathrm{SO}(4))$ generates the so-called  modified L\'evy Laplacian $\Delta_L^W$.
The Lie group $\mathrm{SO}(4)$ is not simple. There are two normal Lie subgroups $S^3_L$ and $S^3_R$ of $\mathrm{SO}(4)$.
The Lie algebra $\mathrm{so}(4)$ can be decomposed as a direct sum $\mathrm{so}(4)=\mathrm{Lie}(S^3_L)\oplus \mathrm{Lie}(S^3_R)$, where $\mathrm{Lie}(S^3_L)$ and $\mathrm{Lie}(S^3_R)$ are the Lie algebras of the groups $S^3_L$ and $S^3_R$ respectively. This decomposition 
corresponds to instantons and antiinstantons. The following theorem was proved for instantons
over $\mathbb{R}^4$  by the author in~\cite{VolkovLLI}.

\begin{theorem}
Let $A$ be a smooth connection in the vector bundle over $R^4$ with fiber $\mathbb{C}^N$.
 Let the value of the Yang--Mills action functional 
  is finite on a connection $A$:
  $$
  -\int_{\mathbb{R}^4}\mathrm{tr}(F_{\mu\nu}(x)F^{\mu\nu}(x))dx<\infty.
  $$
Let $U^A$ be the  parallel transport   generated by the connection $A$.
Let $W\in C^1([0,1],S^3_R)$ ($W\in C^1([0,1],S^3_L)$) and
$$\mathrm{dim}\, \mathrm{span} \{W^{-1}(t)\dot{W}(t)\}_{t\in[0,1]}\geq2.$$
 The following two assertions are equivalent:
\begin{enumerate}
\item  the connection $A$ on $\mathbb{R}^4$ is  an antiinstanton (instanton);
\item  the parallel transport  $U^A$ is a solution of  the  Laplace  equation for the modified L\'evy Laplacian $\Delta_L^{W}$:
\begin{equation*}
 \Delta_L^{W} U^A=0.
\end{equation*}
\end{enumerate}
\end{theorem}
In the present paper, we generalize this theorem for instantons and antiinstantons on a 4-dimensional Riemannian manifold.
The effect of the holonomy group on a manifold on the connection between L\'evy Laplacians and instantons was studied in~\cite{Volkov2022}.

The concept of the modified L\'evy Laplacian can be generalized to the case of a manifold.
It is well-known fact that if $f$ is a smooth  function on a $d$-dimensional Riemannian manifold $M$ than the value of the
Laplace--Beltrami operator $\Delta_M$ on $f$ at a point $x\in M$ can be found as
\begin{equation}
\label{Lap-Beltr}
\Delta_M f(x)=\sum_{i=1}^d \left.\frac{d^2}{dt^2}\right|_{t=0} f(\mathrm{exp}_x{(tZ_i)}),
\end{equation}
where $\mathrm{exp}_x$ is an exponential mapping on the manifold $M$ at the point $x\in M$
and $\{Z_1,\ldots,Z_d\}$ is an orthonormal basis in the tangent space $T_xM$.
In~\cite{AS2006} by Accardi and Smolyanov, the L\'evy Laplacian was introduced on the infinite dimensional 
manifold by analogy with formulas~(\ref{Levy2}) and~(\ref{Lap-Beltr}). In the present paper, we generalize this definition for
modified L\'evy Laplacians.

Different approaches to the Yang--Mills equations  based on the parallel transport but not based on the L\'evy Laplacian
were used in~\cite{Gross,Driver,Bauer1998,Bauer2003,ABT,AT}. Theory of infinite dimensional Laplacians on infinite dimensional manifolds was also considered in~\cite{BP2016,BP2017}. For a recent development in the study of the L\'evy Laplacian in the white noise theory see~\cite{Volkov2018,AHJS}. For a recent development in the study of infinite dimensional Laplacians see~\cite{Sa,BuSa,ZaSa}.

The paper is organized as follows. In Sec.~\ref{Sec1}, we give preliminary information about infinite dimensional geometry. In Sec.~\ref{Sec2}, we give the definition of the L\'evy Laplacian.  
In Sec.~\ref{Sec3}, we   give preliminary information on the Yang--Mills equations and instantons.   We 
formulate the main theorem on  relationship between instantons on a 4-dimensional manifold and instantons  in~Sec.~\ref{Sec4}  and prove it in Sec.~\ref{Sec5}.

\section{Hilbert manifold of $H^1$-curves}
\label{Sec1}

In this section, we give some background information about the Hilbert manifold of $H^1$-curves and two canonical Hilbert vector bundles 
$\mathcal H^1\subset \mathcal H^0$ over this manifold. This is an analog of the embedding  $H^1_0([0,1],\mathbb{R}^d)\subset L_2([0,1],\mathbb{R}^d)$
 from the definition of the L\'evy Laplacian~(\ref{Levy2}).

Let $M$ be a smooth connected (not necessary compact) Riemannian $d$-dimensional manifold. Let $g$ denote the Riemannian metric on $M$.
Let $\Gamma^\kappa_{\lambda\nu}$  be the Christoffel symbols of the Levi-Civita connection associated with this metric on $M$.
We will raise and lower indices using the metric $g$  and we will sum over repeated indices.

For any sub-interval $I\subset[0,1]$ the symbols
$H^0(I,\mathbb{R}^d)$  and $H^1(I,\mathbb{R}^d)$   denote the spaces of $L_2$-functions
and $H^1$-functions  on
$I$ with values in $\mathbb{R}^d$ respectively.  These spaces are  Hilbert spaces with scalar products
$$(h_1,h_2)_0=\int_I(h_1(t),h_2(t))_{\mathbb{R}^d}dt$$
and
   $$(h_1,h_2)_1=\int_I(h_1(t),h_2(t))_{\mathbb{R}^d}dt+\int_I(\dot{h}_1(t),\dot{h}_2(t))_{\mathbb{R}^d}dt$$
   respectively.
Let $H_0^1=\{h\in H^1([0,1],\mathbb{R}^d)\colon h(0)=0)$
and $H_{0,0}^1=\{h\in H_0^1\colon h(1)=0\}$.

The curve $\gamma\colon [0,1]\to M$  on the manifold $M$ is called $H^1$-curve, if $\phi_a\circ \gamma\mid_I\in H^1(I,\mathbb{R}^4)$
 for any interval  $I\subset [0,1]$ and for any  coordinate chart
 $(\phi_a, W_a)$ of the manifold  $M$ such that  $\gamma(I)\subset W_a$.
 Let $\Omega_m$ be the set of all $H^1$-curves in $M$ with the origin at the point $m\in M$.
The  set $\Omega_m$  can be endowed with the structure of an infinite dimensional Hilbert manifold modeled over the Hilbert space $H^1_0$ (see~\cite{Driver,Klingenberg,Klingenberg2,Volkov2019a}).

For any $\gamma\in \Omega_m$ the mapping  $X(\gamma;\cdot)\colon [0,1]\to TM$ such that   $X(\gamma;t)\in T_{\gamma(t)}M$ for any $t\in[0,1]$ is a vector field along  $\gamma$. We will also use the notation $X(\gamma)$ for $X(\gamma;\cdot)$.

The symbol  $H^0_\gamma(TM)$ denotes the Hilbert space of all $H^0$-fields along  $\gamma$. The scalar product
on this space is defined by the formula
$$
G_{0}(X(\gamma),Y(\gamma))=\int_0^1g(X(\gamma;t),Y(\gamma;t))dt.
$$

For any absolutely continuous field  $X(\gamma)$  along  $\gamma\in \Omega$
its covariant derivative $\nabla X(\gamma)$ is defined by
$$
\nabla X(\gamma;t)= \frac {d}{dt}X(\gamma;t)+\Gamma(\gamma(t))(X(\gamma;t),\dot{\gamma}(t)),
$$
where $(\Gamma(x)(X,Y))^\mu=\Gamma^\mu_{\lambda \nu}(x)X^\lambda Y^\nu$ in local coordinates.
Let $Q(\gamma;\cdot)$ denote the parallel transport generated by the Levi-Civita connection  along the curve  $\gamma$.
It is easy to show that
\begin{equation}
\label{covder}
\nabla{X}(\gamma;t)=Q(\gamma;t)\frac d{dt}(Q(\gamma;t)^{-1}X(\gamma;t)).
\end{equation}
The symbol
$H^1_\gamma(TM)$ denotes the Hilbert space of all $H^1$-fields along  $\gamma$. The scalar product
on this space is defined by the formula
$$
G_1(X(\gamma),Y(\gamma))=\int_0^1g(X(\gamma;t),Y(\gamma;t))dt
\\+\int_0^1g(\nabla X(\gamma;t),\nabla Y(\gamma;t))dt.
$$

We consider two canonical Riemannian vector bundles $\mathcal H^0$ and $\mathcal H^1$  over the Hilbert manifold $\Omega_m$ (see~\cite{Klingenberg,Klingenberg2}).
The fiber of $\mathcal H^0$ over  $\gamma\in \Omega_m$ is the space   $H^0_\gamma(TM)$.
The vector bundle   $\mathcal H^1$ is the tangent bundle over the manifold   $\Omega_m$.
Its fiber  over $\gamma\in \Omega_m$ is the space  $H^1_\gamma(TM)$.  
The exponential mapping  at $\gamma\in \Omega_m$
is defined by the formula
\begin{equation}
\mathrm{exp}_\gamma(X)(t)=\mathrm{exp}_{\gamma(t)}(X(t)), \text{ where $X\in H^1_\gamma(TM)$}.
\end{equation}

We can identify $\mathbb{R}^d$ and   the  Hilbert spaces $H^1([0,1],\mathbb{R}^d)$ and  with $T_mM$ and $H^1([0,1],T_mM)$ respectively. 
 Let  fix  an orthonormal basis $\{Z_1,\ldots,Z_d\}$ in $T_mM$.
 Let $Z_i(\gamma,t)=Q_{t,0}(\gamma)Z_i$ for $i\in \{1,\ldots,d\}$.
   Due to~(\ref{covder}), for any $\gamma\in \Omega_m$ the Levi-Civita connection generates the canonical isometrical isomorphism  between
$H^1_0$ and $H^1_\gamma(TM)$, which action on $h\in H^1_0$ we will denote by $\widetilde{h}$. This isomorphism acts   by the formula
\begin{equation}
\label{isomorhism}
\widetilde{h}(\gamma;t)=Q_{t,0}(\gamma)h(t)=Z_\mu(\gamma,t)h^\mu(t),
\end{equation}
where $h=h^\mu Z_\mu$.

\section{Definition of modified L\'evy Laplacian on manifold} 
\label{Sec2}

In this section, we give a definition of the L\'evy Laplacian on the space of functions
on the Hilbert manifold of $H^1$-curves as the Cesaro mean of the second order directional derivatives.

 Everywhere below we assume that $\{e_n\}$ is  an orthonormal basis in $L_2([0,1],\mathbb{R})$ such that
 $e_n\in H^1$ and $e_n(0)=e_n(1)=0$ for all $n\in \mathbb{N}$.
Let $e_{\mu, n}=Z_\mu e_n$. Then $\{e_{\mu, n}\}$ form the orthonormal basis in $H^0$  and all its elements belong
to $H^1_{0,0}$. Hence,
$\{\widetilde{e_{\mu, n}}(\gamma)\}$ form the orthonormal basis in  the fiber of $\mathcal H^0$ over  $\gamma\in \Omega_m$
which elements belong to the fiber of $\mathcal H^1$.

Any smooth curve $W\in C^1([0,1],SO(d))$ defines an orthogonal operator 
in $L_2([0,1],\mathbb{R}^d)$ by pointwise left multiplication:
$$
(Wu)(t)=W(t)u(t).
$$
The subspace  $H^1_0([0,1],\mathbb{R}^d)\subset L_2([0,1],\mathbb{R}^d)$ is invariant under the action of $W$.

Let $\mathfrak{F}(\Omega_m,\mathbb{R})$ denote the space of all functions on $\Omega_m$. 

\begin{definition}
\label{def1}  The (modified) L\'evy Laplacian, generalized by the orthonormal basis $\{e_{\mu,n}\}$  and the curve $W\in C^1([0,1],SO(d))$, is a linear mapping $$\Delta^{W,\{e_{\mu,n}\}}_L:\mathrm{dom}\Delta^{W,\{e_{\mu,n}\}}_L\to
\mathfrak{F}(\Omega_m,\mathbb{R})$$ defined by
\begin{equation}
\label{formld}
 \Delta^{W,\{e_{\mu,n}\}}_Lf(\gamma)=\lim_{n\to\infty}\frac 1n\sum_{k=1}^n\sum_{\mu=1}^d\left.\frac {d^2}{ds^2}\right|_{s=0}f(\mathrm{exp}(s W(\widetilde{e_{\mu,k}}(\gamma))),
\end{equation}
where  $\mathrm{dom} \Delta^{W,\{e_{\mu,n}\}}_L$ is the space of all functions $f\in\mathfrak{F}(\Omega_m,\mathbb{R})$ such that the right side of~(\ref{formld}) exists for all $\gamma\in \Omega_m$.  
\end{definition}

\begin{remark} 
This definition is a  generalization of the definition of the  L\'evy Laplacian on a manifold  introduced by Accardi and Smolyanov in~\cite{AS2006} and the definition of the  L\'evy Laplacian  generalized by the curve $W\in C^1([0,1],SO(4))$, introduced by the author for  $M=\mathbb{R}^4$ in~\cite{VolkovLLI}.
\end{remark}

Recall the following definition from~\cite{L1951} (see also~\cite{Feller}).

\begin{definition} An orthonormal basis $\{e_n\}$ in $L_2([0,1],\mathbb{R})$
is called  weakly uniformly dense (or equally dense) if 
\begin{equation}
\label{unifdence}
\lim_{n\to\infty}\int_0^1 h(t)\left(\frac 1n\sum_{k=1}^n
e_k^2(t)-1\right)dt=0
\end{equation}
for any $h\in L_{\infty}([0,1],\mathbb{R})$.
\end{definition}

\begin{example}
The basis 
 $e_n(t)=\sqrt{2}\sin{(n\pi t)}$ is a weakly uniformly dense basis.~\cite{L1951,Feller}
\end{example}

The following example belongs to 	
Polishchuk.~\cite{Feller}
\begin{example}
Let $r\in C([0,1],\mathbb{R})$. Let $u_n$ be the normalized $n$-th eigenfunction of the following second order differential 
equation:
$$
u''(t)-r(t)u(t)+\lambda u(t)=0
$$
with boundary condition: $u(0)=u(1)=0$. Then $\{u_n\}_{n=1}^\infty$ is an equally dense orthonormal basis in $L_2([0,1],\mathbb{R})$. 
\end{example} 
 
One can show that if $\{e_n\}$ is  weakly uniformly dense basis than the L\'evy Laplacian $\Delta_L^{W,\{e_{\mu,n}\}}$ for constant $W$ coincides with  the L\'evy Laplacian from Ref.~\cite{LV2001} and, in the general case $W\in C^1([0,1],SO(d))$, coincides with the modified L\'evy Laplacian from Ref.~\cite{Volkov2020}.
 
 \begin{example}
\label{example1}
Let $\mathfrak{f}\in C^2(M,\mathbb{R})$. Let
$\mathfrak{L_{f}}\colon \Omega_m\to \mathbb{R}$
be defined by:
$$
\mathfrak{L_{f}}(\gamma)=\int_0^1\mathfrak{f}(\gamma(t))dt.
$$
Then for an arbitrary $W\in C^1([0,1],SO(d))$ the value of the L\'evy Laplacian  on the functional $\mathfrak{L_{f}}$ is
$$
 \Delta^{W}_L \mathfrak{L_{f}}(\gamma)=\int_0^1 \Delta_{(M,g)} \mathfrak{f}(\gamma(t))dt,
$$
where $\Delta_{(M,g)}$ is the Laplace-Beltrami operator on the manifold $M$.
\end{example}
 
\begin{remark}
If $W$ from the definition of the L\'evy Laplacian  $\Delta^{W}_L$ is constant then  we obtain the L\'evy Laplacian which is related to the Yang--Mills equations (see~\cite{AGV1994,LV2001,Volkov2012,Volkov2021}).
\end{remark}

\section{Yang--Mills equations and instantons}
\label{Sec3}

In this section, $M$ is a smooth orientable (not necessary compact) Riemannian $4$-dimensional manifold ($d=4$).

Let $G\subseteq  SU(N)$ be a closed Lie group and $\mathrm{Lie}(G)\subseteq su(N)$ be its Lie algebra.
Let $E=E(\mathbb{C}^N,\pi,M,G)$  be a vector bundle over $M$ with the projection $\pi\colon E\to M$, the fiber $\mathbb{C}^N$ 
and the structure group $G$. We will denote the fiber  $\pi^{-1}(x)\cong\mathbb{C}^N$ over $x\in M$ by the symbol $E_x$. Let $P$ be the principle bundle over $M$  associated with $E$.
 Let $\mathrm{ad} (P)=\mathrm{Lie}(G)\times_G M$ be the adjoint  bundle of $P$  (the fiber of $\mathrm{ad}P$ is
  isomorphic to $\mathrm{Lie} (G)$).

A connection $A(x)=A_\mu(x)dx^\mu$ in the vector bundle $E$ is  a smooth section in
$\Lambda^1(T^\ast M)\otimes \mathrm{ad}P$.  
The curvature $F(x)=\sum_{\mu<\nu}F_{\mu\nu}(x)dx^\mu\wedge dx^\nu$ of the connection $A$ is a smooth section in $\Lambda^2(T^\ast M)\otimes \mathrm{ad}P$ defined by the formula $F_{\mu \nu}=\partial_\mu A_\nu-\partial_\nu A_\mu+[A_\mu,A_\nu]$.

The Yang--Mills action functional has the form
\begin{equation}
\label{YMaction1}
S_{YM}(A)=-\frac 12\int_{M}\mathrm{tr}(F_{\mu\nu}(x)F^{\mu\nu}(x))Vol(dx),
\end{equation}
where $Vol$ is the Riemannian volume measure on the manifold $M$.
The Yang--Mills equations  on a connection $A$ have the form
\begin{equation}
\label{YMequations}
D_A^\ast F=0,
\end{equation}
where $D_A$ is the operator of the exterior covariant derivative and $D_A^\ast$ is its formally adjoint operator. 
In local coordinates, we have
$$
D_A^\ast F_\nu=-\nabla^\mu F_{\mu\nu}
$$
and
 \begin{equation*}
 \nabla_\lambda F_{\mu
\nu}=\partial_\lambda F_{\mu \nu}+[A_\lambda,F_{\mu
\nu}]-F_{\mu
\kappa}\Gamma^\kappa_{\lambda\nu}-F_{\kappa\nu}\Gamma^\kappa_{\lambda\mu},
\end{equation*}
where $\Gamma^\kappa_{\lambda\nu}$  are the Christoffel symbols of the Levi-Civita connection on $M$.
The Yang--Mills equations are the Euler-Lagrange equations for the Yang--Mills action functional~(\ref{YMaction1}).

If $M$ is a 4-manifold, it is possible to consider 
the Yang--Mills self-duality equations 
\begin{equation}
\label{dualintro}
F_{-}:=\frac 12(F-\ast F)=0
\end{equation} or  the Yang--Mills  anti self-duality   equations 
\begin{equation}
\label{antidualintro}
F_{+}:=\frac 12(F+\ast F)=0
\end{equation}
 on a connection $A$, where $\ast$  is the Hodge star.
A connection
 is called  an instanton or  an anti-instanton if it is a solution of equations~(\ref{antidualintro}) or~(\ref{dualintro}) respectively.  Due to the Bianchi identities $D_AF=0$ and equality  $D_A^{\ast}=-\ast D_A\ast$,  instantons and
anti-instantons are solutions of the Yang--Mills equations~(\ref{YMequations}).

\section{Laplace equation and instantons}
\label{Sec4}

Let  $\mathcal{E}_m$ be a Hilbert vector bundle over the Hilbert manifold $\Omega_m$ such that its fiber over  $\gamma\in \Omega_m$ is the space  $\mathrm{Hom}(E_m,E_{\gamma(1)})$. The definition of the modified L\'evy Laplacian  $\Delta_L^{W,\{e_{\mu,n}\}}$  can been transferred  without changes to the space of sections in the bundle $\mathcal{E}_m$.

 For any  $H^1$-curve  $\gamma$  an operator $U^A_{t,s}(\gamma)\in \mathrm{Hom}(E_{\gamma(s)},E_{\gamma(t)})$, where $0\leq s\leq t \leq1$, is a solution  of the system 
 \begin{equation}
\label{form 2}
 \left\{
\begin{aligned}
\frac
{d}{dt}U^A_{t,s}(\gamma)=-A_{\mu}(\gamma(t))\dot{\gamma}^{\mu}(t) U^A_{t,s}(\gamma)\\
\frac
d{ds}U^A_{t,s}(\gamma)=U^A_{t,s}(\gamma)A_{\mu}(\gamma(s))\dot{\gamma}^{\mu}(s)\\
\left.U^A_{t,s}(\gamma)\right|_{t=s}=I_N.
\end{aligned}
\right.
\end{equation}
Then $U^A_{1,0}$ is a parallel transport along the curve $\gamma$ generated by the connection $A$.

The parallel transport has the following properties (see~\cite{Volkov2020}):
\begin{enumerate}
\item The mapping $\Omega_m\ni \gamma\mapsto U^A_{1,0}(\gamma)$ is a $C^{\infty}$-smooth section in the vector bundle $\mathcal E_m$ (for the proof of smoothness see~\cite{Gross,Driver}).
\item The parallel transport does not depend on the choice of parametrization of the curve.
Let $\sigma\colon[0,1]\to [0,1]$ be a
non-decreasing piecewise $C^1$-smooth function such that $\sigma(0)=0$ and $\sigma(1)=1$. 
Then
\begin{equation}
\label{reparametr}
U^A_{\sigma(t),\sigma(s)}(\gamma)=U^A_{t,s}(\gamma\circ\sigma)
\end{equation}
for any $\gamma\in \Omega_m$.
\item For any $\gamma \in \Omega_m$ the  parallel transport satisfies the multiplicative property:
\begin{equation}
\label{group}
U^A_{t,s}(\gamma)U^A_{s,r}(\gamma)=U^A_{t,r}(\gamma)\text{ for $r\leq s\leq t$}.
\end{equation}
\item If the restriction of $\gamma\in \Omega_m$ on $[s,t]$ is constant, then
\begin{equation}
\label{indent}
U^A_{t,s}(\gamma)\equiv I_N.
\end{equation}
\end{enumerate}

 The dimension $d=4$ is special, because where two normal subgroups $S^3_L$ and $S^3_R$ of $SO(4)$. The groups $S^3_L$ and $S^3_R$ consist of real matrices of the form
$$\begin{pmatrix}
a&-b&-c&-d\\
b&\;\,\, a&-d&\;\,\, c\\
c&\;\,\, d&\;\,\, a&-b\\
d&-c&\;\,\, b&\;\,\, a
\end{pmatrix}\text{ and }
\begin{pmatrix}
a&-b&-c&-d\\
b&\;\,\, a&\;\,\, d&-c\\
c&-d&\;\,\, a&\;\,\, b\\
d&\;\,\, c&-b&\;\,\, a
\end{pmatrix},
$$
where $a^2+b^2+c^2+d^2=1$, respectively. 
 The Lie algebras  $\mathrm{Lie}(S^3_L)$ and $\mathrm{Lie}(S^3_R)$ of the Lie groups  $S^3_L$ and $S^3_R$ consist of real  matrices of the form   
$$\begin{pmatrix}
0&-b&-c&-d\\
b&\;\,\, 0&-d&\;\,\, c\\
c&\;\,\, d&\;\,\, 0&-b\\
d&-c&\;\,\, b&\;\,\, 0
\end{pmatrix}\text{ and }
\begin{pmatrix}
0&-b&-c&-d\\
b&\;\,\, 0&\;\,\, d&-c\\
c&-d&\;\,\, 0&\;\,\, b\\
d&\;\,\, c&-b&\;\,\, 0
\end{pmatrix}
$$
respectively.

If $W\in C^1([0,1],\mathrm{SO}(4))$ let $L_W(t)=W^{-1}(t)\dot{W}(t)$. Then $L_W$ is continuous 
curve in  $\mathrm{so}(4)$. Let $L^+_W(t)$ and $L^-_W(t)$ denote the orthogonal projection $L_W(t)$ on the Lie algebras  $\mathrm{Lie}(S^3_L)$ and $\mathrm{Lie}(S^3_R)$ respectively. Let us show that the operator $\Delta_L^{W,\{e_{\mu,n}\}}$ is related to instantons and  antiinstantons,  if $W\in C^1([0,1],\mathrm{Lie}(S^3_L))$ or $W\in C^1([0,1],\mathrm{Lie}(S^3_R))$ respectively (see also~\cite{VolkovLLI,Volkov2020,Volkov2021,Volkov2022}).
Let us prove the following theorem.

\begin{theorem}
Let $W\in C^1([0,1],S^3_L)$ ($W\in C^1([0,1],S^3_R)$) and
$$\mathrm{dim}\,\mathrm{span} \{L_W(t)\}_{t\in[0,1]}\geq2.$$
Let $\{e_n\}$ be  a weakly uniformly dense orthonormal basis in $L_2([0,1],\mathbb{R})$ such that
 $e_n\in H^1$ and $e_n(0)=e_n(1)=0$ for all $n\in \mathbb{N}$.
Let there exists the sequence $\{x_n\}$ of points  in $M$ such that 
\begin{equation}
\label{condition}
\lim\limits_{n\to \infty}F_+(x_n)=0 \;(\lim\limits_{n\to \infty}F_-(x_n)=0).
\end{equation}
 The following two assertions are equivalent:
\begin{enumerate}
\item  a connection $A$ is a solution of the anti self-duality equations (self-duality equations): $F=-\ast F$, ($F=\ast F$);
\item  the parallel transport  $U^A_{1,0}$ is a solution of the equation:
\begin{equation*}
 \Delta_L^{W,\{e_{\mu,n}\}} U^A_{1,0}=0.
\end{equation*}
\end{enumerate}
\end{theorem}

\begin{remark}
If the base manifold $M$ is compact than condition~(\ref{condition}) means that
there exists  a point $x\in M$ such that $F_+(x)=0$ ($F_-(x)=0$).  
\end{remark}

\section{Proof of the main theorem}
\label{Sec5}

In this section, we will prove several lemmas, which together provide a proof of the main theorem.

We will use the following notations for   all $\gamma\in \Omega_m$ and $t\in[0,1]$:
\begin{equation}
L(\gamma,t)=U^A_{t,0}(\gamma)^{-1}F(\gamma(t))U^A_{t,0}(\gamma).
\end{equation}
Its anti self-dual and self-dual parts have form
\begin{equation}
L_{-}(\gamma,t)=U^A_{t,0}(\gamma)^{-1}F_{-}(\gamma(t))U^A_{t,0}(\gamma),
\end{equation}
\begin{equation}
L_{+}(\gamma,t)=U^A_{t,0}(\gamma)^{-1}F_{+}(\gamma(t))U^A_{t,0}(\gamma).
\end{equation}
If $\gamma\in \Omega_m$ and  $r,t\in[0,1]$ let
$$
L^W(\gamma,t,r):=\sum_{\mu=1}^4L(\gamma,t)<\dot{W}(r)e_\mu(\gamma,t),W(r)e_\mu(\gamma,t)>.
$$
It can be checked by direct computations that
\begin{multline*}
L^W(\gamma,t,r)=\mathrm{tr}_{so(4)}(\dot{W}(r)W^{-1}(r)L(\gamma,t))=\\=\mathrm{tr}_{so(4)}({L}_W^+(r)L_+(\gamma,t))+\mathrm{tr}_{so(4)}({L}_W^-(r)L_-(\gamma,t)).
\end{multline*}

The parallel transport belongs to the domain of the modified L\'evy Laplacian.
The following theorem can be proved by direct computations.
\begin{lemma}
\label{lemma1}
Let $\{e_n\}$ be  a weakly uniformly dense orthonormal basis in $L_2([0,1],\mathbb{R})$ such that
 $e_n\in H^1$ and $e_n(0)=e_n(1)=0$ for all $n\in \mathbb{N}$.
The value of the modified  L\'evy Laplacian $\Delta_L^{W,\{e_{\mu,n}\}}$ on  the  parallel transport is
\begin{multline}
\label{lllaplace0}
\Delta_L^{W,\{e_{\mu,n}\}} U^A_{1,0}(\gamma)=\\=\int_0^1U^A_{1,t}(\gamma) D_A^\ast F(\gamma(t))\dot{\gamma}(t)U^A_{t,0}(\gamma)dt
-U^A_{1,0}(\gamma) \int_0^1L^W(\gamma,t,t)dt.
\end{multline}
\end{lemma}

The first term in the right side of~(\ref{lllaplace0}) is invariant under reparameterization of the curve $\gamma$
but the second  term is not. Therefore, the following lemma is true (for the proof see~\cite{Volkov2020}).
\begin{lemma}
\label{lemma2}
Let $W\in C^1([0,1],SO(4))$. Let the parallel transport  $U^A_{1,0}$ be s a solution of the Laplace equation for the L\'evy Laplacian   $\Delta_L^{W,\{e_{\mu,n}\}}$:
$$
 \Delta_L^{W,\{e_{\mu,n}\}} U^A_{1,0}=0.
$$
 Then the connection $A$ satisfies the Yang--Mills equations
   and for any $\gamma\in \Omega_m$ the following holds
  $$
 \int_0^1 L^W(\gamma,t,t)dt=0.
  $$
\end{lemma}

The following lemma is an analog for manifolds of Proposition~3 from Ref.~\cite{VolkovLLI}.
\begin{lemma}
\label{lemma3}
Let
for any $\gamma\in \Omega_m$ the following relation holds
  $$
 \int_0^1 L^W(\gamma,t,t)dt=0.
  $$
Then for any $\gamma\in \Omega_m$ and for any $t,r\in [0,1]$
the following holds
$$
L^W(\gamma,t,r)=\mathrm{tr}_{so(4)}(L_W(r)F(m))).
$$
\end{lemma}
\textbf{Proof}
For any $\gamma\in \Omega_m$ and for any $t\in [0,1]$ and $r\in [0,1]$ let the curve $\gamma\in \Omega_m$ be introduced by the following:
$$\gamma_{r,\varepsilon}(\tau)=\begin{cases}
m, &\text{if $0\leq \tau\leq r-\varepsilon$,}\\
\gamma(\frac {\tau-r+\varepsilon}{\varepsilon}), &\text{if $r-\varepsilon <\tau \leq r$,}\\
\gamma(t),  &\text{if $r<\tau\leq1$.}\\
\end{cases}
$$
We can use the properties of the parallel transport.
If $r<\tau\leq1$, then
$$
L^W(\gamma_{r,\varepsilon},\tau,r)=\mathrm{tr}_{so(4)}(L_W(r)U^A_{t,0}(\gamma)^{-1}F(\gamma(t))U^A_{t,0}(\gamma))=L^W(\gamma,t,r).
$$
If $0\leq \tau\leq r-\varepsilon$, then
$$
L^W(\gamma_{r,\varepsilon},\tau,r)=\mathrm{tr}_{so(4)}(L_W(r)U^A_{t,0}(\gamma)^{-1}F(\gamma(t))U^A_{t,0}(\gamma))=\mathrm{tr}_{so(4)}(L_W(r)F(m)).
$$
Hence,
\begin{multline*}
\int_0^1L^W(\gamma_{r,\varepsilon},\tau,\tau)d\tau=\\=\int_{r}^1L^W(\gamma,t,\tau)d\tau+
\int_{r-\varepsilon}^{r}L^W(\gamma_{r,\varepsilon},\tau,\tau)d\tau+\int_0^{r-\varepsilon}\mathrm{tr}_{so(4)}(L_W(\tau)F(m))d\tau=0.
\end{multline*}

There exists $C>0$ such that the following estimates hold
$$
\left\|\int_{r-\varepsilon}^{r}L^W(\gamma_{r,\varepsilon},\tau,\tau)d\tau\right\|_{su(N)}\leq C\varepsilon,
$$
$$
\left\|\int_{r-\varepsilon}^r\mathrm{tr}_{so(4)}(L_W(\tau)F(m))d\tau\right\|_{su(N)}\leq C\varepsilon.
$$
Then for any $\varepsilon>0$ 
\begin{multline*}
\|\int_{r}^1L^W(\gamma,t,\tau)d\tau+\int_0^{r}\mathrm{tr}_{so(4)}(L_W(\tau)F(m))d\tau\|_{su(N)}=\\
=\|\int_{r}^1L^W(\gamma,t,\tau)d\tau+\int_0^{r}\mathrm{tr}_{so(4)}(L_W(\tau)F(m))d\tau-\int_0^1L^W(\gamma_{r,\varepsilon},\tau,\tau)d \tau
\|_{su(N)}\leq\\
\leq\|\int_{r-\varepsilon}^r\mathrm{tr}_{so(4)}(L_W(\tau)F(m))d\tau-\int_{r-\varepsilon}^{r}L^W(\gamma_{r,\varepsilon},\tau,\tau)d\tau\|_{su(N)}\leq 2C\varepsilon.
\end{multline*}
Then
$$
-\int_{r}^1L^W(\gamma,t,\tau)d\tau=\int_0^{r}\mathrm{tr}_{so(4)}(L_W(\tau)F(m)))d\tau.
$$
Differentiating the last equality we get
$$
L^W(\gamma,t,r)=\mathrm{tr}_{so(4)}(L_W(r)F(m)).
$$

\begin{lemma}
\label{lemma4}
Let $W\in C^1([0,1],S^3_L)$  ($W\in C^1([0,1],S^3_R)$).
Let
for any $\gamma\in \Omega_m$ the following holds
  $$
 \int_0^1 L^W(\gamma,t,t)dt=0.
  $$
Let there exists the sequence $\{x_n\}$ in $M$ such that 
\begin{equation}
\label{condition1}
\lim\limits_{n\to \infty}F_+(x_n)=0\; (\lim\limits_{n\to \infty}F_-(x_n)=0).
\end{equation}
Then for any $\gamma\in \Omega_m$ and for any $t,r\in [0,1]$
the following holds
$$
L^W(\gamma,t,r)=0.
$$
\end{lemma}
\textbf{Proof} 
If $W\in C^1([0,1],S^3_L)$, then ${L}_W^-(r)=0$ and
$$
L^W(\gamma,t,r)=\mathrm{tr}_{so(4)}(\dot{W}(r)W^{-1}(r)L(\gamma,t))=\mathrm{tr}_{so(4)}({L}_W^+(r)L_+(\gamma,t)).
$$
Let there exists the sequence $\{x_n\}$ in $M$ such that 
$$
\lim\limits_{n\to \infty}F_+(x_n)=0,
$$
then for any $r\in[0,1]$ there exist  sequences of curves $\{\gamma_n\}$ and points $\{t_n\}$ such that
$x_n=\gamma_n(t_n)$. Due to Lemma~\ref{lemma3}   for any $\gamma\in \Omega_m$ and for any $t,r\in [0,1]$ we have
$$
L^W(\gamma,t,r)=\mathrm{tr}_{so(4)}(L_W(r)F(m))=\lim_{n\to \infty}L^W(\gamma_n,t_n,r)=0.
$$

\begin{lemma}
\label{lemma5}
Let a connection $A$ satisfy the  Yang--Mills equations  
\begin{equation}
%\label{YMequations}
D_A^\ast F=0.
\end{equation}
Let $W\in C^1([0,1],S^3_L)$ ($W\in C^1([0,1],S^3_R)$) and
\begin{equation}
\label{ineq}
\mathrm{dim}\,\mathrm{span} \{L_W(t)\}_{t\in[0,1]}\geq2.
\end{equation}
Let  for any $\gamma\in \Omega_m$ and for any $t,r\in [0,1]$
the following holds
$$
L^W(\gamma,t,r)=0.
$$
Then the  connection $A$ is a solution of self--duality equations (anti self--duality equations): $F=\ast F$, ($F=-\ast F$).
\end{lemma}
\textbf{Proof}
Lemma~\ref{lemma4} implies that  for any $\gamma\in \Omega_m$ and for any $t,r\in [0,1]$
the following holds
$L^W(\gamma,t,r)=0.$
This equality and  inequality~(\ref{ineq}) together
imply that we can choose the orthonormal basis $\{Z_1,Z_2,Z_3,Z_4\}$ in $T_mM$ such that for any $t\in[0,1]$
\begin{equation}
\label{eqq1}
 \left\{
\begin{aligned}
L_+(\gamma,t)\langle Z_1(\gamma,t)\wedge Z_2(\gamma,t)\rangle=0\\
 L_+(\gamma,t)\langle Z_1(\gamma,t)\wedge Z_3(\gamma,t)\rangle=0.
\end{aligned}
\right.
\end{equation}
Differentiating~(\ref{eqq1}) and multiplying left and right by $U^A_{t,0}(\gamma)$ and $\left(U^A_{t,0}(\gamma)\right)^{-1}$, respectively we get
\begin{equation}
\label{nablaf+1}
\nabla F_{+}(\gamma(t))\langle \dot{\gamma}(t),Z_1(\gamma,t),Z_2(\gamma,t)\rangle=0
\end{equation}
and
\begin{equation}
\label{nablaf+2}
\nabla F_{+}(\gamma(t))\langle \dot{\gamma}(t),Z_1(\gamma,t),Z_3(\gamma,t)\rangle=0.
\end{equation}

The Yang--Mills equations imply that
\begin{multline}
\label{YM22}
0=\nabla F(\gamma(t))\langle Z_1(\gamma,t),Z_1(\gamma,t), Z_4(\gamma,t)\rangle+\\+\nabla F(\gamma(t))\langle Z_2(\gamma,t),Z_2(\gamma,t), Z_4(\gamma,t)\rangle+\\ +\nabla F(\gamma(t)) \langle  Z_3(\gamma,t),Z_3(\gamma,t), Z_4(\gamma,t)  \rangle.
\end{multline}
The Bianchi identities imply that
\begin{multline}
\label{Bianchi22}
0= \nabla F(\gamma(t)) \langle Z_1(\gamma,t),Z_3(\gamma,t),Z_2(\gamma,t)\rangle+\\+\nabla F(\gamma(t))\langle Z_2(\gamma,t),Z_1(\gamma,t),Z_3(\gamma,t)\rangle+\\+\nabla F(\gamma(t))\langle Z_3(\gamma,t),Z_2(\gamma,t), Z_1(\gamma,t)\rangle.
\end{multline}
Equalities~(\ref{YM22}) and~(\ref{Bianchi22}) together imply
\begin{multline}
\label{nablaFF}
\nabla F(\gamma(t)) \langle Z_1(\gamma,t),Z_1(\gamma,t),Z_4(\gamma,t)\rangle-\nabla F(\gamma(t)) \langle Z_1(\gamma,t),Z_3(\gamma,t),Z_2(\gamma,t)\rangle=\\=\nabla F(\gamma(t)) \langle Z_2(\gamma,t),Z_1(\gamma,t),Z_3(\gamma,t)\rangle-\\-\nabla F(\gamma(t)) \langle Z_2(\gamma,t),Z_2(\gamma,t),Z_4(\gamma,t)\rangle+\\+\nabla F(\gamma(t))\langle Z_3(\gamma,t), Z_2(\gamma,t),Z_1(\gamma,t)\rangle-\\-\nabla F(\gamma(t))\langle Z_3(\gamma,t), Z_3(\gamma,t),Z_4(\gamma,t)\rangle.
\end{multline}

Equalities~(\ref{nablaFF}) can be rewritten in the form
\begin{multline}
\label{nablaFF1}
\nabla F_+(\gamma(t))\langle Z_1(\gamma,t),Z_1(\gamma,t),Z_2(\gamma,t)\rangle=\\=\nabla F_+(\gamma(t))\langle Z_1(\gamma,t),Z_3(\gamma,t),Z_4(\gamma,t)\rangle=\\
=\nabla F_+(\gamma(t))\langle Z_2(\gamma,t),Z_1(\gamma,t),Z_3(\gamma,t)\rangle+\\+\nabla F_+(\gamma(t))\langle  Z_3(\gamma,t), Z_2(\gamma,t),Z_1(\gamma,t) \rangle.
\end{multline}

Due to~(\ref{nablaf+1}) and to~(\ref{nablaf+2}), the right side of~(\ref{nablaFF1}) is zero.
Hence,
\begin{multline*}
\nabla F_+(\gamma(t))\langle Z_1(\gamma,t),Z_1(\gamma,t),Z_2(\gamma,t)\rangle=\\=\nabla F_+(\gamma(t))\langle Z_1(\gamma,t),Z_3(\gamma,t),Z_4(\gamma,t)\rangle=0.
\end{multline*}

By analogous reasoning, one can prove that $$\nabla F_+(\gamma(t))\langle Z_\mu(\gamma,t),Z_\nu(\gamma,t),Z_\lambda(\gamma,t)\rangle=0$$ for all 
$\mu,\nu,\lambda \in \{1,2,3,4\}$.
Hence the tensor $F_+$ is parallel and $\|F_+\|$ is constant on the manifold $M$.
There is a sequence $\{x_n\}$ of points in $M$,
such that $\lim_{n\to\infty}\|F_{+}(x_n)\|=0$.
 Then $F_{+}\equiv0$ and  $A$ is an instanton.

These lemmas together imply the assertion of the main theorem.

\section*{Acknowledgments}
This work was supported by the Russian Science Foundation under grant no. 19-11-00320, https://rscf.ru/en/project/19-11-00320/

\bibliographystyle{unsrt}
\bibliography{Volkovarxiv}

\end{document}